\documentclass[lettersize,journal]{IEEEtran}

\usepackage{amsmath,amssymb,amsfonts}
\usepackage{algorithmic}
\usepackage{algorithm}
\usepackage{array}
\usepackage{textcomp}
\usepackage{stfloats}
\usepackage{url}
\usepackage{verbatim}
\usepackage{graphicx}
\usepackage{cite}
\def\BibTeX{{\rm B\kern-.05em{\sc i\kern-.025em b}\kern-.08em
		T\kern-.1667em\lower.7ex\hbox{E}\kern-.125emX}}
\usepackage{balance}

\usepackage{subfigure}
\usepackage{epstopdf}

\makeatletter
\def\endthebibliography{%
	\def\@noitemerr{\@latex@warning{Empty `thebibliography' environment}}%
	\endlist
}
\makeatother

\begin{document}

\include{header}

\title{Exploiting Moving Arrays for Near-Field Sensing}

\author{Yilong Chen, Zixiang Ren, Xianghao Yu, Lei Liu, and Jie Xu
	\thanks{Y. Chen, Z. Ren, and J. Xu are with the School of Science and Engineering (SSE), the Shenzhen Future Network of Intelligence Institute (FNii-Shenzhen), and the Guangdong Provincial Key Laboratory of Future Networks of Intelligence, The Chinese University of Hong Kong, Shenzhen, Guangdong 518172, China (e-mail: yilongchen@link.cuhk.edu.cn; rzx66@mail.ustc.edu.cn; xujie@cuhk.edu.cn).}
	\thanks{Z. Ren is also with the Key Laboratory of Wireless-Optical Communications, Chinese Academy of Sciences, School of Information Science and Technology, University of Science and Technology of China, Hefei 230027, China.}
	\thanks{X. Yu is with the Department of Electrical Engineering, City University of Hong Kong, Hong Kong, China (e-mail: alex.yu@cityu.edu.hk).}
	\thanks{L. Liu is with Anhui Jiaoxin Technology Co., Ltd, Hefei 230041, China (e-mail: liul313@foxmail.com).}
	\thanks{J. Xu is the corresponding author.}
}

\maketitle

\begin{abstract}
	This letter exploits moving arrays to enable near-field multiple-input multiple-output (MIMO) sensing via a limited number of antenna elements. We consider a scenario where a base station (BS) is equipped with a uniform linear array (ULA) on a moving platform. The objective is to locate a point target in the two-dimensional (2D) space by leveraging the near-field channel characteristics created by the movement of antenna arrays. Under this setup, we analyze the Cram{\'e}r-Rao bound (CRB) for estimating the target’s 2D coordinate, which provides the fundamental sensing performance limits for localization. It is revealed that our proposed design with a moving array achieves a CRB that is proportional to the CRB obtained by an equivalent extremely large ULA matching the platform's size. This shows that the movement of antenna array significantly enlarges its effective aperture to enable near-field sensing. Numerical results show that the proposed moving array design substantially enhances the target estimation performance compared to the conventional fixed array benchmark.
\end{abstract}

\vspace{-0.1cm}

\vspace{-0.1cm}\section{Introduction}

The integration of sensing has been widely recognized as a key feature of sixth-generation (6G) wireless networks \cite{liu2022integrated, chen2024isac}. Among other sensing techniques, multiple-input multiple-output (MIMO) sensing is particularly appealing to enhance the sensing accuracy and resolution, in which multiple antennas are deployed at base station (BS) transceivers to provide array gains and spatial diversity gains. More recently, with the advancements in extremely large antenna arrays (ELAA), hundreds to thousands of antennas are equipped at the BS to substantially increase the aperture and enable near-field sensing for beam focusing and target localization \cite{wang2024cramer, hua2023near}. However, the extremely large number of antenna elements in ELAA may introduce significant fabrication costs and energy consumption, thus hindering its wide application in practice.

Recently, moving arrays \cite{edelson1992ability, sheinvald1998localization, dogandzic2001cramer, boyer2011performance, yang2019direction, qin2019doa, patra2022novel, luo2023angle} have attracted growing interest as an emerging MIMO radar technique, in which the antenna array is deployed on a moving platform to move collectively, thus creating a large virtual aperture to enhance the sensing performance with low cost.\footnote{Notice that there is another line of research on movable antenna \cite{zhu2024MA, ma2024movable} or fluid antenna \cite{wong2020performance}, which allows to reconfigure each antenna's position of the array individually for reshaping MIMO channels to facilitate sensing and communications. However, movable/fluid antennas may stay at the optimized positions over a certain time period for sensing or communications. Differently, our considered moving array can move continuously over time to enhance the effective aperture for enabling near-field sensing.}
There have been a handful of prior works studying moving array enabled sensing from different perspectives. For instance, the authors in \cite{edelson1992ability} investigated the Cram{\'e}r-Rao bound (CRB) for direction of arrival (DoA) estimation when the array moves at a constant speed, and \cite{sheinvald1998localization} proposed a DoA estimator based on generalized least squares (GLS). Furthermore, the authors in \cite{dogandzic2001cramer, boyer2011performance, yang2019direction} studied the estimation of both DoA and velocity of moving targets via exploiting co-located MIMO moving arrays. More recently, the authors in \cite{qin2019doa, patra2022novel, luo2023angle} proposed to deploy sparse antenna arrays on the moving platform to further increase the degrees of freedom (DoFs) for DoA estimation. However, these prior works on moving array enabled sensing focused on the case of far-field sensing, by assuming the planar wavefront models.

Different from prior works, this letter is the first attempt to exploit moving arrays to enable near-field sensing by deploying a relatively small antenna array on a large-scale moving platform. For initial investigation, we study a basic scenario with a BS equipped with a co-located uniform linear array (ULA) on the moving platform, which aims to locate a point target in the two-dimensional (2D) space. We consider the near-field channel property created by the array movement, in which exact spherical wavefront models are employed \cite{hua2023near}.
To reveal the fundamental performance limits of such systems, we analyze the CRB for estimating the target’s 2D coordinate.
We show that with the moving array, the estimation CRB is highly dependent on the transmit signal waveform over time. This is in sharp contrast to the conventional fixed array enabled sensing, in which the CRB only depends on the sample covariance matrix. We also show that the estimation CRB achieved by the moving array is proportional to that by an equivalent extremely large ULA matching the platform's size. This indicates that the movement of antenna array significantly enlarges its effective size and enables near-field sensing for locating a distant target, which is infeasible for conventional fixed arrays with far-field sensing. Finally, numerical results are provided to validate the 2D localization performance of our proposed moving array enabled near-field sensing.

\textit{Notations:} For an arbitrary-size matrix \(\boldsymbol{A}\), \(\boldsymbol{A}^*\), \(\boldsymbol{A}^T\), and \(\boldsymbol{A}^H\) denote its conjugate, transpose, and conjugate transpose, respectively. For a square matrix \(\boldsymbol{A}\), \(\mathrm{tr}(\boldsymbol{A})\) and \(\boldsymbol{A}^{-1}\) denote its trace and inverse, respectively. For a vector \(\boldsymbol{a}\), \(\|\boldsymbol{a}\|\) denotes its Euclidean norm. For a complex number \(a\), \(\Re\{a\}\) and \(\Im\{a\}\) denote its real and imaginary parts, respectively. \(\mathbb{C}^{M \times N}\) denotes the space of \(M \times N\) complex matrices.

\vspace{-0.1cm}\section{System Model}

\begin{figure}[tb]
	\centering {\includegraphics[width=0.32\textwidth]{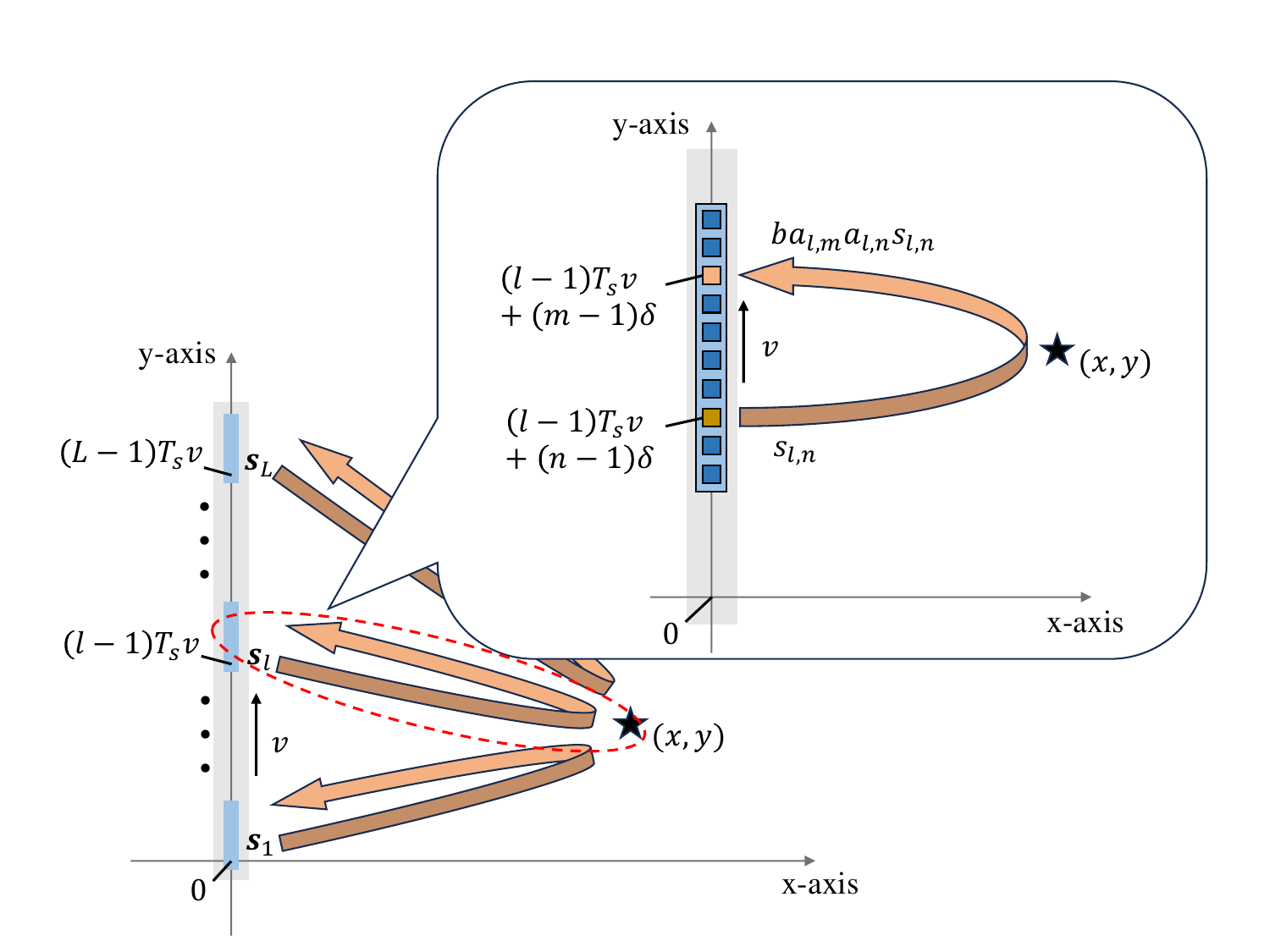}} 
	\caption{Moving array enabled near-field target localization.}
	\label{Fig_sys}
\vspace{-0.1cm}\end{figure}
As shown in Fig. \ref{Fig_sys}, we consider a wireless sensing scenario, in which a BS is equipped with a ULA on a moving platform, aiming to locate a point target at coordinate \((x,y)\) in the 2D space. We assume that there is a set \(\mathcal{N} = \{1, \dots, N\}\) of \(N\) co-located antenna elements at the ULA, with the half-wavelength spacing between adjacent antennas denoted by \(\delta\). Without loss of generality, we suppose that the platform is deployed on the y-axis and the ULA moves along the y-axis in the positive direction at a given speed \(v\).\footnote{The antenna array can move back in the opposite direction in the next round.}

We focus on the sensing operation over a finite duration \(T\), which consists of a set \(\mathcal{L} = \{1,\dots,L\}\) of \(L\) symbols each with duration \(T_s\), i.e., \(T=LT_s\). Consequently, the platform's size is \(LT_s v\).
Let \(\boldsymbol{s}_l = [{s}_{l,1}, \dots, {s}_{l,N}]^T\) denote the transmit signal vector by the ULA-BS at symbol \(l \in \mathcal{L}\). Accordingly, the sample covariance matrix is given by \(\boldsymbol{R} = \frac{1}{L} \sum_{l=1}^L \boldsymbol{s}_l \boldsymbol{s}_l^H\), which should satisfy the transmit power constraint \(\mathrm{tr}(\boldsymbol{R}) \le P_0\), with \(P_0\) denoting the maximum transmit power.

It is assumed that at symbol \(l=1\), antenna \(n=1\) of the ULA is located at the origin \((0,0)\). Thus, at each symbol \(l\in\mathcal{L}\), antenna \(n\in\mathcal{N}\) of the ULA is located at coordinate \((0,\bar{y}_{l,n})\), with \(\bar{y}_{l,n} = (l-1)T_sv + (n-1)\delta\), such that the distance between antenna \(n\) and the target is \(d_{l,n} = \sqrt{x^2 + y_{l,n}^2}\), with \(y_{l,n} = y - \bar{y}_{l,n}\).
Accordingly, we define \(\boldsymbol{a}_l = [a_{l,1}, \dots, a_{l,N}]^T\) as the steering vector of the ULA at symbol \(l\), where \(a_{l,n}\) is given by \cite{hua2023near}
\vspace{-0.2cm}\begin{equation} \label{a_l}
	a_{l,n} = \frac{\lambda}{4\pi d_{l,n}} e^{-j\frac{2\pi}{\lambda}d_{l,n}}.
\vspace{-0.2cm}\end{equation}
In \eqref{a_l}, the exact spherical wavefront model with respect to (w.r.t.) the carrier wavelength \(\lambda\) \cite{hua2023near} is utilized to capture the near-field effect, which is different from conventional sensing designs based on far-field channel models, e.g., \cite{edelson1992ability, sheinvald1998localization, dogandzic2001cramer, boyer2011performance, yang2019direction, qin2019doa, patra2022novel, luo2023angle}.

The received signal at each symbol \(l\in\mathcal{L}\) by the ULA-BS is expressed as\footnote{In \eqref{y_l}, we neglect the array movement during the signal propagation. This is reasonable as the round-trip signal transmission delay over the BS transmitter-target-BS receiver link is much shorter than symbol duration \(T_s\) in practice. Furthermore, we assume that the Doppler shifts induced by the array movement can be perfectly compensated via proper signal processing, as the antenna array moves at a preset speed with pre-known Doppler shifts.
}
\vspace{-0.2cm}\begin{equation} \label{y_l}
	\boldsymbol{r}_l = b \boldsymbol{a}_l \boldsymbol{a}_l^T \boldsymbol{s}_l + \boldsymbol{z}_l.
\vspace{-0.2cm}\end{equation}
In \eqref{y_l}, \(\boldsymbol{z}_l \in\mathbb{C}^{N\times1}\) denotes the noise at the ULA-BS, which is a circularly symmetric complex Gaussian (CSCG) random vector with zero mean and covariance \(\sigma^2 \boldsymbol{I}_N\), i.e., \(\boldsymbol{z}_l \sim \mathcal{CN}(\boldsymbol{0}_N, \sigma^2 \boldsymbol{I}_N)\). 
Furthermore, \(b\) denotes the complex target reflection coefficient, which is proportional to the radar-cross-sections (RCS) of the isotropic target \cite{wang2024cramer, hua2023near}.
To facilitate target localization, we combine the received signals in \eqref{y_l} as \(\boldsymbol{r} = [\boldsymbol{r}_1^T, \dots, \boldsymbol{r}_L^T]^T\).
The objective of target localization is to estimate unknown parameters  \(\boldsymbol{\eta} = \big[x, y, \Re\{b\}, \Im\{b\}\big]^T\) based on observation \(\boldsymbol{r} \sim \mathcal{CN}\big(\boldsymbol{\mu}(\boldsymbol{\eta}), \boldsymbol{\Sigma}(\boldsymbol{\eta})\big)\), with \(\boldsymbol{\mu}(\boldsymbol{\eta}) = \big[(b \boldsymbol{a}_1 \boldsymbol{a}_1^T \boldsymbol{s}_1)^T, \dots, (b \boldsymbol{a}_L \boldsymbol{a}_L^T \boldsymbol{s}_L)^T\big]^T\) and \(\boldsymbol{\Sigma}(\boldsymbol{\eta}) = \sigma^2 \boldsymbol{I}_{NL}\). 

\vspace{-0.1cm}\section{Near-field Estimation CRB}

In this section, we derive the CRB for target localization achieved by our proposed design with moving array, which serves as the lower bound of the variance of any unbiased estimators.\footnote{We will show the practical near-field target localization performance in Section V by considering a maximum likelihood estimator.}

Let \(\boldsymbol{F}\in\mathbb{C}^{4\times4}\) denote the Fisher information matrix (FIM) w.r.t. the parameters \(\boldsymbol{\eta}\). Based on the observation \(\boldsymbol{r}\), the \((i,j)\)-th element of \(\boldsymbol{F}\) is given by \cite{hua2023near}
\vspace{-0.15cm}\begin{equation} \label{F_ij}
	\begin{aligned}
		\boldsymbol{F}(\eta_i, \eta_j) &= \mathrm{tr}\left\{\boldsymbol{\Sigma}^{-1}(\boldsymbol{\eta}) \frac{\partial \boldsymbol{\Sigma}(\boldsymbol{\eta})}{\partial \eta_i} \boldsymbol{\Sigma}^{-1}(\boldsymbol{\eta}) \frac{\partial \boldsymbol{\Sigma}(\boldsymbol{\eta})}{\partial \eta_j}\right\} \\
		&+ 2\Re \mathrm{tr} \left\{\big(\frac{\partial \boldsymbol{\mu}(\boldsymbol{\eta})}{\partial \eta_i}\big)^H \boldsymbol{\Sigma}^{-1}(\boldsymbol{\eta}) \frac{\partial \boldsymbol{\mu}(\boldsymbol{\eta})}{\partial \eta_j}\right\},
	\end{aligned}
\vspace{-0.15cm}\end{equation}
where \(\eta_i\) and \(\eta_j\) denote the \(i\)-th and \(j\)-th elements of \(\boldsymbol{\eta}\), respectively. We then have the following lemma.

\textbf{Lemma 1}: The FIM for estimating \(\boldsymbol{\eta}\) based on \(\boldsymbol{r}\) is given by
\vspace{-0.35cm}\begin{equation} \label{F}
	\boldsymbol{F} = 2 \begin{bmatrix}
		\Re\{F_{xx}\} & \Re\{F_{xy}\} & \Re\{F_{xb}\} & -\Im\{F_{xb}\} \\
		\Re\{F_{xy}\} & \Re\{F_{yy}\} & \Re\{F_{yb}\} & -\Im\{F_{yb}\} \\
		\Re\{F_{xb}\} & \Re\{F_{yb}\} & \Re\{F_{bb}\} & 0 \\
		-\Im\{F_{xb}\} & -\Im\{F_{yb}\} & 0 & \Re\{F_{bb}\}
	\end{bmatrix},
\vspace{-0.2cm}\end{equation}
where \(F_{pq} = \frac{|b|^2}{\sigma^2} \sum_{l=1}^L \boldsymbol{s}_l^H \boldsymbol{\dot{A}}_{p,l}^H \boldsymbol{\dot{A}}_{q,l} \boldsymbol{s}_l\), \(F_{pb} = \frac{b^*}{\sigma^2} \sum_{l=1}^L \boldsymbol{s}_l^H \boldsymbol{\dot{A}}_{p,l}^H \boldsymbol{A}_{l} \boldsymbol{s}_l\), and \(F_{bb} = \frac{1}{\sigma^2} \sum_{l=1}^L \boldsymbol{s}_l^H \boldsymbol{A}_l^H \boldsymbol{A}_l \boldsymbol{s}_l\),
\(\forall p,q \in \{x,y\}\). Here, for notational convenience, we define \(\boldsymbol{A}_l = \boldsymbol{a}_l \boldsymbol{a}_l^T\) and \(\boldsymbol{\dot{A}}_{p,l} = \frac{\partial \boldsymbol{A}_l}{\partial p} = \boldsymbol{\dot{a}}_{p,l} \boldsymbol{a}_l^T + \boldsymbol{a}_l \boldsymbol{\dot{a}}_{p,l}^T\), where \(\boldsymbol{\dot{a}}_{p,l} = \frac{\partial \boldsymbol{a}_l}{\partial p}\) denotes the derivative of \(\boldsymbol{a}_l\) w.r.t. \(p \in \{x,y\}\).

\textit{Proof}: See Appendix A. \hfill \(\square\)

Given the FIM, the CRB matrix w.r.t. \(\boldsymbol{\eta}\) is defined as \(\boldsymbol{D} = \boldsymbol{F}^{-1}\).
Consequently, the CRB for estimating the target's coordinate \((x,y)\) is given by \(\mathrm{CRB}\big(\{\boldsymbol{s}_l\}\big) = [\boldsymbol{D}]_{1,1} + [\boldsymbol{D}]_{2,2}\), for which we have the following proposition.

\textbf{Proposition 1}: With the moving array, the estimation CRB is given by
\vspace{-0.3cm}\begin{equation} \label{CRBs}
	\mathrm{CRB}\big(\{\boldsymbol{s}_l\}\big) = \frac{\sigma^2}{2 |b|^2} \frac{G_{yy} + G_{xx}}{G_{xx} G_{yy} - \Re\{G_{xy}\}^2},
\vspace{-0.15cm}\end{equation}
where \(G_{pq}\), \(\forall p,q \in \{x,y\}\), is defined as 
\vspace{-0.15cm}\begin{equation} \label{G}
	\begin{aligned}
		G_{pq} 
		&= \sum_{l\in\mathcal{L}} \mathrm{tr}(\boldsymbol{\dot{A}}_{p,l}^H \boldsymbol{\dot{A}}_{q,l} \boldsymbol{s}_l \boldsymbol{s}_l^H) \\
		&- \frac{\sum_{l\in\mathcal{L}} \mathrm{tr}(\boldsymbol{\dot{A}}_{p,l}^H \boldsymbol{A}_l \boldsymbol{s}_l \boldsymbol{s}_l^H) \sum_{l=1}^L \mathrm{tr}(\boldsymbol{A}_l^H \boldsymbol{\dot{A}}_{q,l} \boldsymbol{s}_l \boldsymbol{s}_l^H)}{\sum_{l=1}^L \mathrm{tr}(\boldsymbol{A}_l^H \boldsymbol{A}_l \boldsymbol{s}_l \boldsymbol{s}_l^H)}.
	\end{aligned}
\vspace{-0.1cm}\end{equation}

\textit{Proof}: See Appendix B. \hfill \(\square\)

\textbf{Remark 1}: It is observed from Proposition 1 that  \(\mathrm{CRB}\big(\{\boldsymbol{s}_l\}\big)\) is a function of the transmit signal waveform \(\boldsymbol{s}_l\)'s over different symbols. This is in sharp contrast to the conventional design with fixed arrays, in which the estimation CRB only depends on the sample covariance matrix \cite{hua2023near}. In this context, the signal waveform design is crucial for the localization performance of the moving array enabled sensing. 

\vspace{-0.1cm}\section{CRB Comparison with Fixed Arrays}

This section compares the estimation CRB performance achieved by the moving array in \eqref{CRBs}, versus those by the following two fixed arrays.

\begin{itemize}
	\item \textbf{Conventional fixed array with \(N\) antennas}: This corresponds to fixing the moving array at the starting point, i.e., each antenna \(n\in\mathcal{N}\) is located at \(\bar{y}_{1,n}\). In this case, the array size is \(N\delta\), and the corresponding estimation CRB is given by \cite{hua2023near}
	\vspace{-0.25cm}\begin{equation} \label{CRBl}
		\overline{\mathrm{CRB}}(\boldsymbol{R}) = \frac{\sigma^2}{2 |b|^2 L} \frac{\bar{G}_{yy} + \bar{G}_{xx}}{\bar{G}_{xx} \bar{G}_{yy} - \Re\{\bar{G}_{xy}\}^2},
		\vspace{-0.2cm}\end{equation}
	where \(\bar{G}_{pq}\), \(\forall p,q \in \{x,y\}\), is defined as
	\vspace{-0.2cm}\begin{equation} \label{G_l}
		\begin{aligned}
			\bar{G}_{pq} 
			&= \mathrm{tr}(\boldsymbol{\dot{A}}_{p,1}^H \boldsymbol{\dot{A}}_{q,1} \boldsymbol{R}) - \frac{\mathrm{tr}(\boldsymbol{\dot{A}}_{p,1}^H \boldsymbol{A}_1 \boldsymbol{R}) \mathrm{tr}(\boldsymbol{A}_1^H \boldsymbol{\dot{A}}_{q,1} \boldsymbol{R})}{\mathrm{tr}(\boldsymbol{A}_1^H \boldsymbol{A}_1 \boldsymbol{R})}.
		\end{aligned}
	\vspace{-0.1cm}\end{equation}
	It is observed from \eqref{CRBl} that \(\overline{\mathrm{CRB}}(\boldsymbol{R})\) only depends on the sample covariance matrix \(\boldsymbol{R}\), instead of the signal waveform \(\boldsymbol{s}_l\)'s in \eqref{CRBs} with moving arrays.
	
	\item \textbf{Extended fixed array with \(L\) antennas}: This corresponds to deploying each antenna \(l\in\mathcal{L}\) at the positions of antenna \(n=1\) of the moving array over different symbols, i.e., \(\{\bar{y}_{l,1}\}\). The size of such an array is \(LT_sv\), matching that of the platform. For this setup, the estimation CRB, denoted by \(\widehat{\mathrm{CRB}}(\hat{\boldsymbol{R}})\), can be expressed similarly as in \eqref{CRBl} by revising \(\boldsymbol{a}_1\) and \(\boldsymbol{R}\) as the steering vector and sample covariance matrix for the constructed \(L\)-dimensional ULA, denoted by \(\hat{\boldsymbol{a}} \in \mathbb{C}^{L\times1}\) and \(\hat{\boldsymbol{R}} \in \mathbb{C}^{L\times L}\), respectively.
\end{itemize}

It is observed that the values of \(\mathrm{CRB}\big(\{\boldsymbol{s}_l\}\big)\), \(\overline{\mathrm{CRB}}(\boldsymbol{R})\), and \(\widehat{\mathrm{CRB}}(\hat{\boldsymbol{R}})\) highly depend on the transmission designs at the ULA-BS. For comparison, we consider the strongest eigenmode (SEM) transmission for the three systems, i.e., we set \(\{\boldsymbol{s}_l^\text{sem} = \frac{\sqrt{P_0}}{\|\boldsymbol{a}_l\|} \boldsymbol{a}_l^*\}\) for the moving array, \(\boldsymbol{R}^\text{sem} = \frac{P_0}{\|\boldsymbol{a}_1\|^2} \boldsymbol{a}_1^* \boldsymbol{a}_1^T\) for the conventional fixed array, and \(\hat{\boldsymbol{R}}^\text{sem} = \frac{P_0}{\|\hat{\boldsymbol{a}}\|^2} \hat{\boldsymbol{a}}^* \hat{\boldsymbol{a}}^T\) for the extended fixed array, respectively.\footnote{The SEM transmission is considered as it is an efficient design for CRB minimization in near-field target localization \cite{hua2024near}.} Also, as the distance between the ULA-BS and target is generally far larger than the size of the moving array \(N\delta\), we can safely neglect the distance variation between the target and different antennas of the moving array, i.e., in the following, we define and assume that \(d_l = d_{l,1} \approx d_{l,n}\) and \(y_l = y_{l,1} \approx y_{l,n}\), \(\forall l\in\mathcal{L}, n\in\mathcal{N}\setminus\{1\}\).

Under the above consideration, we have the approximated CRBs in the following proposition.

\textbf{Proposition 2}: 
Under the SEM transmission and by assuming \(\{d_l \approx d_{l,n}\}\) and \(\{y_l \approx y_{l,n}\}\), the estimation CRBs for the moving array, the conventional fixed array, and the extended fixed array are respectively given by
\vspace{-0.1cm}\begin{align}
	\mathrm{CRB}\big(\{\boldsymbol{s}_l^\text{sem}\}\big) &\approx \frac{\sigma^2}{2 |b|^2 (1-{\alpha})} (\frac{1}{{G}_{xx}^\text{sem}} + \frac{1}{{G}_{yy}^\text{sem}}), \label{CRB2} \\
	\overline{\mathrm{CRB}}(\boldsymbol{R}^\text{sem}) &\approx \frac{\sigma^2}{2 |b|^2 L (1-\bar{\alpha})} (\frac{1}{\bar{G}_{xx}^\text{sem}} + \frac{1}{\bar{G}_{yy}^\text{sem}}), \label{CRBl2} \\
	\widehat{\mathrm{CRB}}(\hat{\boldsymbol{R}}^\text{sem}) &\approx \frac{\sigma^2}{2 |b|^2 L (1-\hat{\alpha})} (\frac{1}{\widehat{G}_{xx}^\text{sem}} + \frac{1}{\widehat{G}_{yy}^\text{sem}}). \label{CRB0}
\vspace{-0.2cm}\end{align}
Here, \(\bar{\alpha} = \frac{\Re\{\bar{G}_{xy}\}^2}{\bar{G}_{xx} \bar{G}_{yy}} <1\) is a parameter indicating the offset of the target w.r.t. the perpendicular bisector of the conventional fixed array. Particularly, if \(\bar{\alpha} = 0\), then it indicates that the offset is zero and the lower bound of \(\overline{\mathrm{CRB}}\) is achieved \cite{hua2023near}. Furthermore, \(\hat{\alpha} <1\) and \({\alpha} = \frac{\Re\{{G}_{xy}\}^2}{{G}_{xx} {G}_{yy}} <1\) are defined similarly as \(\bar{\alpha}\) for the extended fixed array and the moving array, respectively. Additionally, with SEM transmission, the approximations of \(\bar{G}_{pp}\), \(\widehat{G}_{pp}\), and \({G}_{pp}\), \(\forall p\in\{x,y\}\), are respectively given by
\vspace{-0.15cm}\begin{equation} \label{Gxyl}
	\begin{aligned}
		\bar{G}_{xx}^\text{sem} &= \frac{P_0 \lambda^2 x^2}{64\pi^2} \sum_{n=1}^{N} \sum_{m=n+1}^{N} \Big(\frac{1}{d_{1,n}^2 d_{1,m}} - \frac{1}{d_{1,m}^2 d_{1,n}}\Big)^2, \\
		\bar{G}_{yy}^\text{sem} &= \frac{P_0 \lambda^2}{64\pi^2} \sum_{n=1}^{N} \sum_{m=n+1}^{N} \Big(\frac{y_{1,n}}{d_{1,n}^2 d_{1,m}} - \frac{y_{1,m}}{d_{1,m}^2 d_{1,n}}\Big)^2,
	\end{aligned}
\vspace{-0.0cm}\end{equation}
\vspace{-0.1cm}\begin{equation} \label{Gxy0}
	\begin{aligned}
		\widehat{G}_{xx}^\text{sem} &= \frac{P_0 \lambda^2 x^2}{64\pi^2} \sum_{i=1}^{L} \sum_{j=i+1}^{L} \Big(\frac{1}{d_i^2 d_j} - \frac{1}{d_j^2 d_i}\Big)^2, \\
		\widehat{G}_{yy}^\text{sem} &= \frac{P_0 \lambda^2}{64\pi^2} \sum_{i=1}^{L} \sum_{j=i+1}^{L} \Big(\frac{y_i}{d_i^2 d_j} - \frac{y_j}{d_j^2 d_i}\Big)^2,
	\end{aligned}
\vspace{-0.0cm}\end{equation}
\vspace{-0.1cm}\begin{equation} \label{Gxyt}
	\begin{aligned}
		{G}_{xx}^\text{sem} &= \frac{N^2 P_0 \lambda^2 x^2}{16\pi^2} \frac{\sum_{i=1}^{L} \sum_{j=i+1}^{L} \frac{1}{d_i^2 d_j^2} (\frac{1}{d_i^2 d_j} - \frac{1}{d_j^2 d_i})^2}{\sum_{l=1}^{L} \frac{1}{d_l^4}}, \\
		{G}_{yy}^\text{sem} &= \frac{N^2 P_0 \lambda^2}{16\pi^2} \frac{\sum_{i=1}^{L} \sum_{j=i+1}^{L} \frac{1}{d_i^2 d_j^2} (\frac{y_i}{d_i^2 d_j} - \frac{y_j}{d_j^2 d_i})^2}{\sum_{l=1}^{L} \frac{1}{d_l^4}}.
	\end{aligned}
\vspace{-0.0cm}\end{equation}

\textit{Proof}: See Appendix C. \hfill \(\square\)

It is observed from \eqref{Gxy0} and \eqref{Gxyt} that \({G}_{xx}^\text{sem}\) and \({G}_{yy}^\text{sem}\) exhibit similar structures as \(\widehat{G}_{xx}^\text{sem}\) and \(\widehat{G}_{yy}^\text{sem}\), respectively, we thus have the following results. 

\textbf{Proposition 3}: When \(d_l\)'s become sufficiently large or \(\{d_l \to \infty\}\) holds, we have \({G}_{xx}^\text{sem} = \frac{4 N^2}{L^2} \widehat{G}_{xx}^\text{sem}\) and \({G}_{yy}^\text{sem} = \frac{4 N^2}{L^2} \widehat{G}_{yy}^\text{sem}\), and thus
\vspace{-0.25cm}\begin{equation} \label{appr}
	\mathrm{CRB}\big(\{\boldsymbol{s}_l^\text{sem}\}\big) \approx \frac{L^2}{4 N^2} \widehat{\mathrm{CRB}}(\hat{\boldsymbol{R}}^\text{sem}).
\vspace{-0.1cm}\end{equation}

\textbf{Remark 2}: It is observed from Proposition 3 that \(\mathrm{CRB}\big(\{\boldsymbol{s}_l^\text{sem}\}\big)\) for the moving array is approximately proportional to \(\widehat{\mathrm{CRB}}(\hat{\boldsymbol{R}}^\text{sem})\) for the extended fixed array. This result implies that the movement of array enlarges its effective size from \(N\delta\) to \(LT_sv\) and increases its equivalent numbers of antennas from \(N\) to \(L\).

\textbf{Remark 3}:
In Proposition 2, it follows that for the conventional fixed array with a small value of \(N\), we have \(d_{1,n} \approx d_{1}, \forall n\in\mathcal{N}\), and thus \(\bar{G}_{xx}^\text{sem} \to 0\) and \(\bar{G}_{yy}^\text{sem} \to 0\), resulting in \(\overline{\mathrm{CRB}}(\boldsymbol{R}^\text{sem}) \to \infty\). By contrast, for the extended fixed array and the moving array with \(L\) sufficiently large, \(d_l\)'s become different over \(l\in\mathcal{L}\), as the distance to the target is comparable to the platform's size \(LT_sv\). As a result, \(\widehat{G}_{xx}^\text{sem}\) and \(\widehat{G}_{yy}^\text{sem}\), as well as \({G}_{xx}^\text{sem}\) and \({G}_{yy}^\text{sem}\), are generally larger than \(\bar{G}_{xx}^\text{sem}\) and \(\bar{G}_{yy}^\text{sem}\) in orders of magnitude, respectively. Therefore, \(\widehat{\mathrm{CRB}}(\hat{\boldsymbol{R}}^\text{sem})\) and \(\mathrm{CRB}\big(\{\boldsymbol{s}_l^\text{sem}\}\big)\) can remain finite.

Based on Remarks 2 and 3, it follows that while the conventional fixed array fails in locating a distant target due to the far-field channel characteristics, the moving array efficiently enables near-field sensing to make target localization feasible. This will also be shown in numerical results in Section V.

\vspace{-0.1cm}\section{Numerical Results}

This section provides numerical results to verify the target localization performance of our proposed moving array, in comparison to the conventional and extended fixed arrays.
The following system parameters are considered unless specified otherwise.
The target is located at coordinate \((x,y) = (10,0)\) m. The array is equipped with \(N = 16\) antennas with spacing \(\delta = 2.5\) cm, corresponding to the half-wavelength of the carrier wave with a frequency of \(6\) GHz. The array moves at a speed of \(v = 5\) m/s. There are \(L = 1000\) symbols, each with a duration of \(T_s = 1\) ms. 
Consequently, the platform's size, or equivalently the size of the extended fixed array, is \(LT_s v = 5\) m. Under the above setup, the Rayleigh distances for the conventional and extended fixed arrays are given by \(\frac{2(N\delta)^2}{\lambda} = 6.4\) m and \(\frac{2(LT_sv)^2}{\lambda} = 1\) km, respectively \cite{selvan2017fraunhofer}.
Furthermore, the transmit power and the noise power are set to be \(P_0 = 30\) dBm and \(\sigma^2 = -70\) dBm, respectively. 
Besides the SEM transmission, we also consider the isotropic transmission for comparison, i.e., we set \(\boldsymbol{R}^\text{iso} = \frac{P_0}{N} \boldsymbol{I}_N\) for the conventional fixed array, \(\hat{\boldsymbol{R}}^\text{iso} = \frac{P_0}{L} \boldsymbol{I}_L\) for the extended fixed array, and \(\big\{\boldsymbol{s}_l^\text{iso} \sim \mathcal{CN}(\boldsymbol{0}_N, \frac{P_0}{N} \boldsymbol{I}_N)\big\}\) for the moving array, respectively.

\begin{figure*}[tb]
	\centering
	\begin{minipage}{0.32\textwidth}
		{\includegraphics[width=1\textwidth]{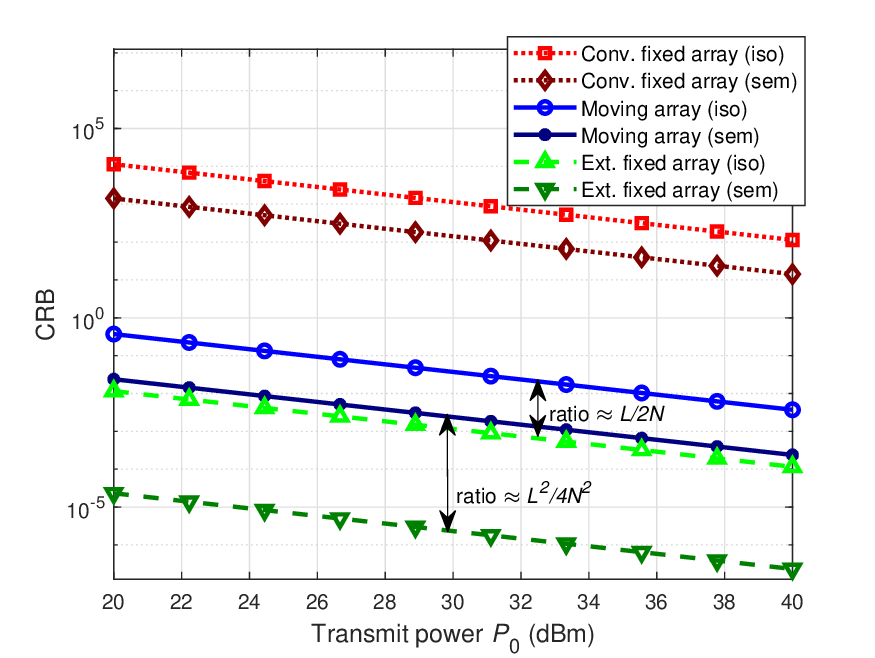}}
		\caption{CRB versus the transmit power \(P_0\).}
		\label{Fig_P}
	\end{minipage}
	\begin{minipage}{0.32\textwidth}
		{\includegraphics[width=1\textwidth]{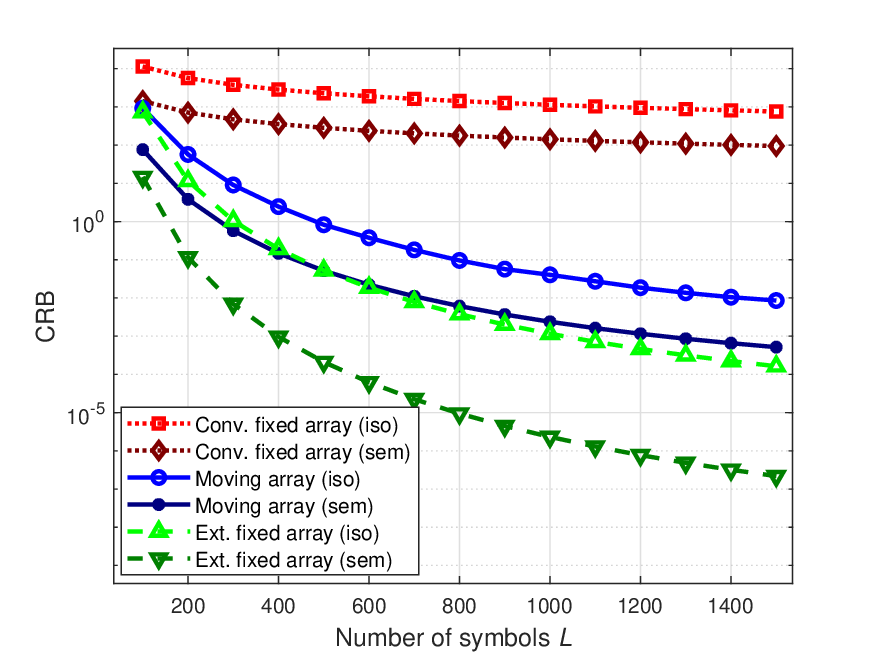}}
		\caption{CRB versus the number of symbols \(L\).}
		\label{Fig_L}
	\end{minipage}
	\begin{minipage}{0.32\textwidth}
		\includegraphics[width=1\textwidth]{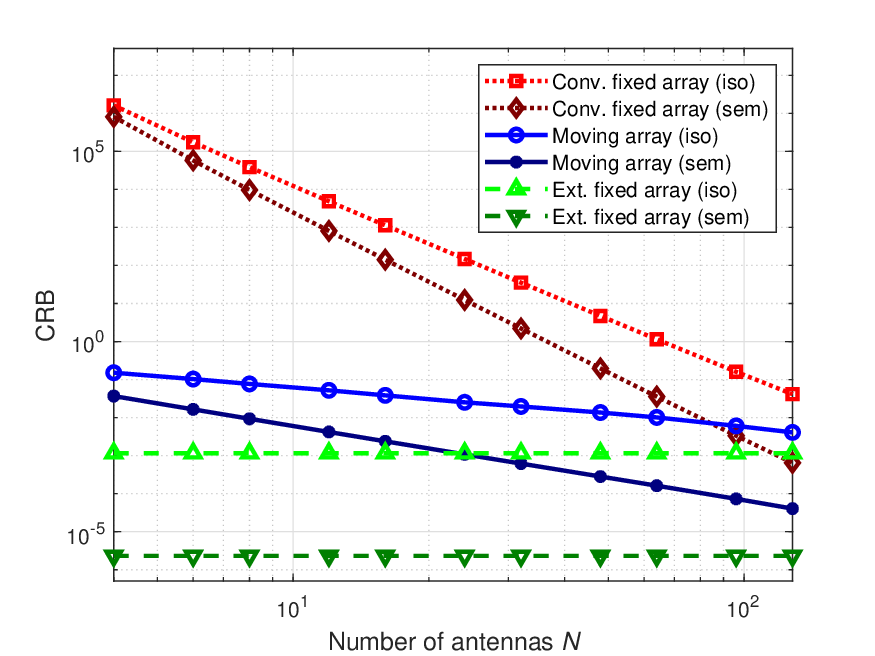}
		\caption{CRB versus the number of antennas \(N\).}
		\label{Fig_N}
	\end{minipage}
\vspace{-0.1cm}\end{figure*}

Fig. \ref{Fig_P} shows the CRB versus the transmit power \(P_0\).
It is observed that under both isotropic and SEM transmissions, the moving array outperforms the conventional fixed array by more than four orders of magnitude. In this case, the localization is only feasible for the moving array and the extended fixed array due to the exploitation of near-field channel characteristics, but infeasible for the conventional fixed array.
In addition, it is observed that the SEM transmission outperforms the isotropic transmission. 
Furthermore, the ratio of CRBs achieved by the moving array and the extended fixed array under the SEM transmission is approximately \(\frac{L^2}{4N}\), which is consistent with the analysis in Proposition 3.


Fig. \ref{Fig_L} shows the CRB versus the number of symbols \(L\).
It is observed that increasing \(L\) enlarges the size of the extended fixed array, or the equivalent size of the moving array, thereby efficiently reducing their CRBs. This is consistent with the analysis in Remark 2. By contrast, increasing \(L\) has a minor effect on the performance of the conventional fixed array.

Fig. \ref{Fig_N} shows the CRB versus the number of antennas \(N\).
It is observed that for the conventional fixed array, increasing \(N\) efficiently enhances its performance. This is consistent with the analysis in Remark 3. In particular, when \(N = 128\), the conventional fixed array achieves a CRB comparable to the moving array with \(N = 16\). By contrast, increasing \(N\) has a minor effect on the performance of the moving array.

\begin{figure}[tb]
	\centering 
	\subfigure[Moving array (maximum log-likelihood at \((10,0)\) m).]
	{\includegraphics[width=0.49\columnwidth]{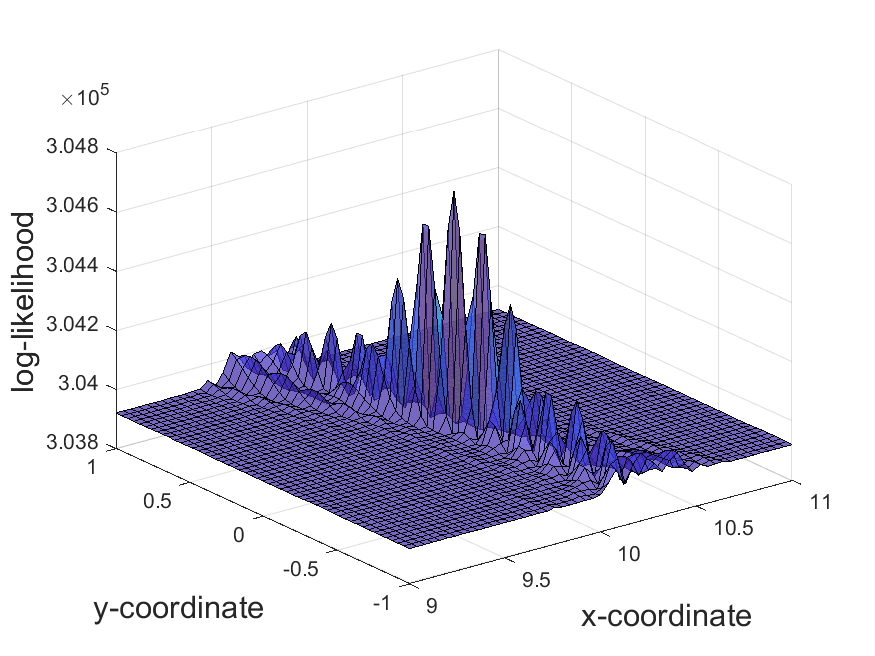}}
	\subfigure[Conventional fixed array (maximum log-likelihood at \((9.88,0.96)\) m).]
	{\includegraphics[width=0.49\columnwidth]{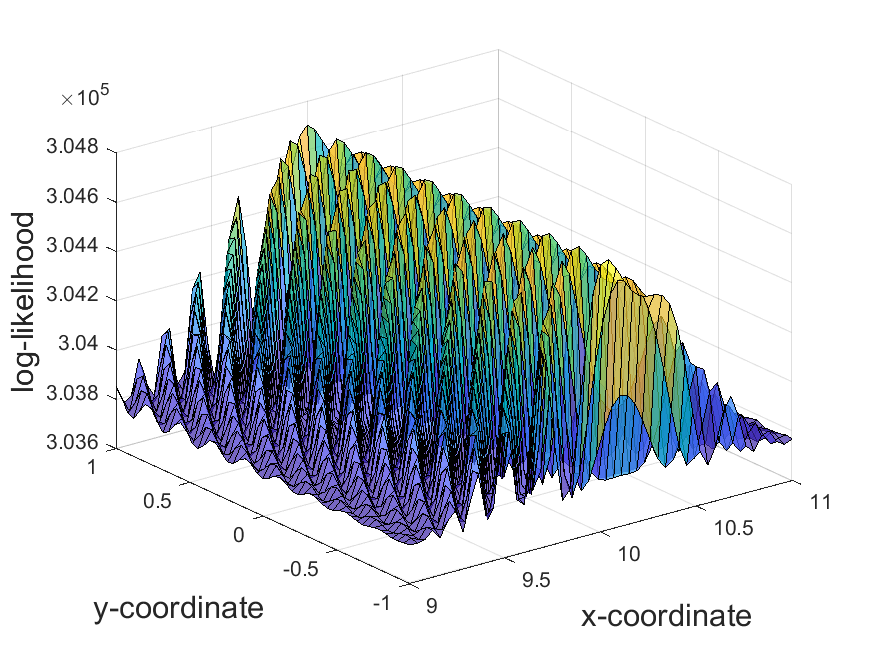}}
	\subfigure[Extended fixed array (maximum log-likelihood at \((10,0)\) m).]
	{\includegraphics[width=0.49\columnwidth]{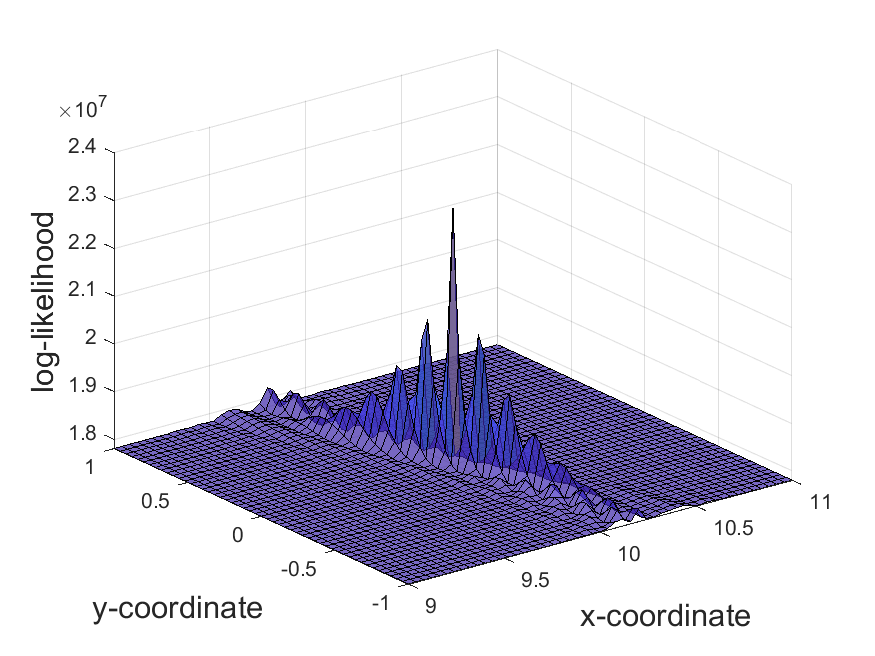}}
	\caption{Log-likelihood function map w.r.t. the 2D coordinate.}
	\label{Fig_lh}
\vspace{-0.1cm}\end{figure}
Finally, we evaluate the practical near-field localization performance, by considering the maximum likelihood estimator building upon the observation \(\boldsymbol{r} \sim \mathcal{CN}\big(\boldsymbol{\mu}(\boldsymbol{\eta}), \boldsymbol{\Sigma}(\boldsymbol{\eta})\big)\). The log-likelihood function of \(\boldsymbol{r}\) is given by
\vspace{-0.2cm}\begin{equation}
	\ln f(\boldsymbol{r}) = \ln \frac{1}{\pi^{LN} |\boldsymbol{\Sigma}(\boldsymbol{\eta})|} - \big(\boldsymbol{r} - \boldsymbol{\mu}(\boldsymbol{\eta})\big)^H \boldsymbol{\Sigma}^{-1}(\boldsymbol{\eta}) \big(\boldsymbol{r} - \boldsymbol{\mu}(\boldsymbol{\eta})\big).
\vspace{-0.2cm}\end{equation}
Maximizing \(\ln f(\boldsymbol{r})\) is equivalent to maximizing \cite{hua2023near}
\vspace{-0.2cm}\begin{equation}
	\tilde{f}(x,y) = -LN\big(\ln\frac{\pi}{LN} + 1 + \ln \sum_{l\in\mathcal{L}} \|\boldsymbol{r}_l - \tilde{b} \boldsymbol{a}_l \boldsymbol{a}_l^T \boldsymbol{s}_l\|^2\big),
\vspace{-0.25cm}\end{equation}
where \(\tilde{b} = \frac{\sum_{l\in\mathcal{L}} \boldsymbol{s}_l^H \boldsymbol{a}_l^* \boldsymbol{a}_l^H \boldsymbol{r}_l}{\sum_{l\in\mathcal{L}} \|\boldsymbol{a}_l \boldsymbol{a}_l^T \boldsymbol{s}_l\|^2}\). Based on this principle, we estimate the target's location by maximizing \(\tilde{f}(x,y)\) via a 2D exhaustive search.
Fig. \ref{Fig_lh} shows the log-likelihood function w.r.t. the 2D coordinate under the SEM transmission. It is observed that the moving array accurately locates the target by achieving a log-likelihood function map with an explicit peak in Fig. \ref{Fig_lh}(a), which is similar to that of the extended fixed array in Fig. \ref{Fig_lh}(c). By contrast, as shown in Fig. \ref{Fig_lh}(b), the conventional fixed array fails to locate the target due to the ambiguity in its log-likelihood function, especially along the y-direction. This validates the effectiveness of the moving array in enabling near-field localization. 

\vspace{-0.1cm}\section{Conclusion}

This letter studied the performance of near-field target localization enabled by the moving array in terms of estimation CRB.
We demonstrated that the CRB achieved by the moving array is proportional to that by the extended fixed array, which implies that the movement of array efficiently increases its equivalent size, thereby enabling near-field sensing and significantly enhancing the estimation performance compared with the conventional fixed array.
Numerical results showed that our proposed moving array achieves a similar estimation performance to those of extended fixed arrays with extremely large scales, thereby significantly reducing manufacturing costs and signaling overhead.

\vspace{-0.1cm}\appendix

\vspace{-0.1cm}\subsection{Proof of Lemma 1}
As \(\boldsymbol{\eta}\)  does not contain any unknowns in \(\boldsymbol{\Sigma}(\boldsymbol{\eta})\), the first term in \eqref{F_ij} vanishes, and we have
\vspace{-0.2cm}\begin{equation} \label{F_ij2}
	\boldsymbol{F}(\eta_i, \eta_j) = 2\Re \{F_{\eta_i \eta_j}\}, \forall \eta_i, \eta_j \in \boldsymbol{\eta},
\vspace{-0.2cm}\end{equation}
where \(F_{\eta_i \eta_j} = \frac{1}{\sigma^2} \sum_{l=1}^L \mathrm{tr} \left\{\big(\frac{\partial (b \boldsymbol{a}_l \boldsymbol{a}_l^T \boldsymbol{s}_l)}{\partial \eta_i}\big)^H \frac{\partial (b \boldsymbol{a}_l \boldsymbol{a}_l^T \boldsymbol{s}_l)}{\partial \eta_j}\right\}\). Here, \(\frac{\partial (b \boldsymbol{a}_l \boldsymbol{a}_l^T \boldsymbol{s}_l)}{\partial \eta_i}\)'s are calculated by
\vspace{-0.2cm}\begin{equation} \label{parti}
	\begin{aligned}
		\frac{\partial (b \boldsymbol{a}_l \boldsymbol{a}_l^T \boldsymbol{s}_l)}{\partial p} &= b \boldsymbol{\dot{a}}_{p,l} \boldsymbol{a}_l^T \boldsymbol{s}_l + b \boldsymbol{a}_l \boldsymbol{\dot{a}}_{p,l}^T \boldsymbol{s}_l, \\
		\frac{\partial (b \boldsymbol{a}_l \boldsymbol{a}_l^T \boldsymbol{s}_l)}{\partial \Re\{b\}} &= -j \frac{\partial (b \boldsymbol{a}_l \boldsymbol{a}_l^T \boldsymbol{s}_l)}{\partial \Im\{b\}} = \boldsymbol{a}_l \boldsymbol{a}_l^T \boldsymbol{s}_l,
	\end{aligned}
\vspace{-0.2cm}\end{equation}
\(\forall p \in \{x,y\}, l \in\mathcal{L}\). Substituting \eqref{parti} into \eqref{F_ij2}, we obtain \eqref{F}, thus completing the proof.

\vspace{-0.1cm}\subsection{Proof of Proposition 1}
According to the Matrix Inversion Lemma, it holds that
\vspace{-0.2cm}\begin{equation}
	\begin{aligned}
	[\boldsymbol{D}]_{1:2,1:2} &= \big([\boldsymbol{F}]_{1:2,1:2} - [\boldsymbol{F}]_{1:2,3:4} [\boldsymbol{F}]_{3:4,3:4}^{-1} [\boldsymbol{F}]_{1:2,3:4}^T\big)^{-1} \\
	&= \left(\frac{2 |b|^2}{\sigma^2} \begin{bmatrix}
		G_{xx} & \Re\{G_{xy}\} \\
		\Re\{G_{xy}\} & G_{yy}
	\end{bmatrix}\right)^{-1}.
	\end{aligned}
\vspace{-0.1cm}\end{equation}
Consequently, we have \(\mathrm{CRB}\big(\{\boldsymbol{s}_l\}\big) = \mathrm{tr}\big([\boldsymbol{D}]_{1:2,1:2}\big)\), thereby leading to \eqref{CRBs} and \eqref{G}. Thus, this completes the proof.

\vspace{-0.1cm}\subsection{Proof of Proposition 2}
For the conventional fixed array, substituting \(\boldsymbol{a}_1\) and \(\boldsymbol{R}^\text{sem}\) into \eqref{G_l}, \(\bar{G}_{xx}\) and \(\bar{G}_{yy}\) are given by
\vspace{-0.2cm}\begin{equation} \label{Gppb}
	\bar{G}_{pp} = P_0 \big(\|\boldsymbol{a}_{1}\|^2 \|\boldsymbol{\dot{a}}_{p,1}\|^2 - |\boldsymbol{a}_{1}^H \boldsymbol{\dot{a}}_{p,1}|^2\big), \forall p \in \{x,y\}.
\vspace{-0.2cm}\end{equation}
Here, it holds that
\vspace{-0.2cm}\begin{equation} \label{v_xy}
	\begin{aligned}
		&\|\boldsymbol{a}_{l}\|^2 = \sum_{n\in\mathcal{N}} \frac{\lambda^2}{16\pi^2 d_{l,n}^2}, \\
		&\|\boldsymbol{\dot{a}}_{x,l}\|^2 \approx \sum_{n\in\mathcal{N}} \frac{x^2}{4 d_{l,n}^4},
		\|\boldsymbol{\dot{a}}_{y,l}\|^2 \approx \sum_{n\in\mathcal{N}} \frac{y_{l,n}^2}{4 d_{l,n}^4}, \\
		&\boldsymbol{\dot{a}}_{x,l}^H \boldsymbol{a}_{l} \approx \sum_{n\in\mathcal{N}} \frac{j \lambda x}{8 \pi d_{l,n}^3},
		\boldsymbol{\dot{a}}_{y,l}^H \boldsymbol{a}_{l} \approx \sum_{n\in\mathcal{N}} \frac{j \lambda y_{l,n}}{8 \pi d_{l,n}^3}, \forall l \in\mathcal{L}.
	\end{aligned}
\vspace{-0.0cm}\end{equation}
Substituting \eqref{v_xy} with \(l=1\) into \eqref{Gppb}, we get the approximations \(\bar{G}_{xx}^\text{sem}\) and \(\bar{G}_{yy}^\text{sem}\) in \eqref{Gxyl}.

For the extended fixed array, \(\widehat{G}_{xx}^\text{sem}\) and \(\widehat{G}_{yy}^\text{sem}\) in \eqref{Gxy0} can be directly obtained via replacing \(\{d_{1,n}\}\) by \(\{d_{l}\}\) and \(\{y_{1,n}\}\) by \(\{y_{l}\}\) in \eqref{Gxyl}, respectively.
 
For the moving array, similar to \eqref{Gppb}, \({G}_{xx}\) and \({G}_{yy}\) are respectively expressed as
\vspace{-0.2cm}\begin{equation} \label{Gppt}
	\begin{aligned}
		{G}_{pp} &= P_0 \sum_{l\in\mathcal{L}} \big(\|\boldsymbol{a}_{l}\|^2 \|\boldsymbol{\dot{a}}_{p,l}\|^2 + 3|\boldsymbol{a}_{l}^H \boldsymbol{\dot{a}}_{p,l}|^2\big) \\
		&- \frac{4P_0 \big(\sum_{l\in\mathcal{L}} \|\boldsymbol{a}_{l}\|^2 \boldsymbol{a}_{l}^H \boldsymbol{\dot{a}}_{p,l}\big) \big(\sum_{l\in\mathcal{L}} \|\boldsymbol{a}_{l}\|^2 \boldsymbol{\dot{a}}_{p,l}^H \boldsymbol{a}_{l}\big)}{\sum_{l\in\mathcal{L}} \|\boldsymbol{a}_{l}\|^4},
	\end{aligned}
\vspace{-0.1cm}\end{equation}
\(\forall p \in \{x,y\}\). Substituting \eqref{v_xy} with approximations \(\{d_l \approx d_{l,n}\}\) and \(\{y_l \approx y_{l,n}\}\) into \eqref{Gppt}, we get the approximations \({G}_{xx}^\text{sem}\) and \({G}_{yy}^\text{sem}\) in \eqref{Gxyt}. Thus, this completes the proof.

\vspace{-0.1cm}
\bibliographystyle{IEEEtran}
\bibliography{MA_ref}

\end{document}